\documentclass[pdflatex,sn-basic]{sn-jnl}

\usepackage{graphicx}
\usepackage{amsmath,amssymb}
\usepackage{booktabs}

\graphicspath{{./}}

\begin{document}

\title[Fractal dimension predicts quantum kernel collapse]{Fractal dimension predicts quantum kernel collapse in angle-encoded data}

\author*[1]{\fnm{Ana Paula} \sur{Appel}}\email{aappel@redhat.com}

\affil*[1]{\orgname{Red Hat}, \orgaddress{\city{S{\~a}o Paulo}, \country{Brazil}}}

\abstract{Angle encoding maps each recorded attribute to one qubit, and
fidelity kernels concentrate as that circuit grows wider.  On a table
with $E$ attributes the encoded width is usually the PCA-95\% cut or
all $E$ columns, chosen so the circuit fits the device.  We propose
the correlation fractal dimension $D_2$ as an \emph{a priori} qubit
budget: encode $\lceil D_2\rceil$ original columns chosen by forward
selection on the partial fractal dimension instead of those cuts.  On nine
data sets and a statevector simulator ($n=32$), a one-layer $ZZ$
fidelity kernel at $q{=}\lceil D_2\rceil$ stays geometrically alive
while the same kernel at the PCA-95\% width has already collapsed.
The budget is map-dependent: product-state and IQP maps overshoot it;
a second $ZZ$ layer undershoots it.  Packed dense-angle and
re-uploading encodings still live at the fractal $q$, but not when
PCA-95\% features are stacked onto those qubits.  Shrinking the
angle bandwidth moves the $ZZ$ knee later; stretching it kills the
kernel earlier.  On IBM Quantum (\texttt{ibm\_fez}, $256$ shots,
$n{=}8$) the one-layer $ZZ$ kernel at the fractal width matches the
exact kernel (MAE $0.021$); past that width both hardware and
simulator have collapsed.  The ceiling is a property of the
map--data pair at a stated bandwidth, not of the classical table
alone.}

\keywords{quantum kernels, intrinsic dimension, fractal dimension, FD-ASE, angle encoding, IBM Quantum}

\maketitle

\section{Introduction}
\label{sec:intro}

Quantum kernel methods on tabular data face a problem that classical
kernels do not: the circuit can collapse before the model has seen any
label.  A tabular data set is a matrix $X\in\mathbb{R}^{N\times E}$
with $N$ records and $E$ measured attributes.  In angle encoding each
attribute becomes a rotation on one qubit, so encoding the full table
spends $E$ qubits.  As the number of qubits grows, the prepared states
$\lvert\psi(x)\rangle$ spread through an exponentially large Hilbert
space, become nearly orthogonal, and the fidelity kernel
\begin{equation}
  \label{eq:kernel}
  K_{ij}=\bigl|\langle\psi(x_i)\mid\psi(x_j)\rangle\bigr|^{2}
\end{equation}
concentrates on the diagonal.  At that point every downstream method
that reads $K$---kernel SVMs, kernel ridge regression, spectral
clustering, one-class novelty detection, nearest-neighbour rules on
fidelities---receives a matrix that is essentially the identity.  It
equally affects variational classifiers whose cost landscape depends
on inner products in Hilbert space.  The model no longer sees which
points are similar and which are not; no training signal can recover
geometry that the encoding has already erased.

This is not a theoretical curiosity.
\citet{thanasilp2024exponential} proved that the concentration is
exponential in the number of qubits for global feature maps.  On
breast cancer data ($E=30$), a $ZZ$ fidelity kernel is geometrically
dead by four qubits.  PCA at $95\%$ variance asks for ten.  The
practitioner who follows the standard recipe---reduce with PCA until
the circuit fits the chip---feeds the quantum model a kernel that has
already collapsed.

The root cause is a mismatch between the circuit width and the data.
The recorded width $E$ is an \emph{embedding} dimension.
The number of degrees of freedom of the support---a curve, a sheet, a
fractal dust---is the \emph{intrinsic} dimension, and it can be far
smaller than $E$.  A circle sitting in $\mathbb{R}^{20}$ still has one
degree of freedom; eighteen axes are padding.  The circuit does not
know that.  Every padded axis is an extra rotation, an extra factor in
the inner product, and another step toward a diagonal $K$.

Current practice already reduces $E$ until the circuit fits the chip:
principal components, autoencoders, or a manual cut.
\citet{belis2025learning} recently proposed learning reduced
representations for quantum classifiers, but the rank $k$ of a PCA
truncation remains a hyperparameter, typically chosen by explained
variance (often $95\%$) or by trying $4$, $8$, $16$ qubits and
looking at accuracy \emph{after} the circuit has run.  That order is
backwards if the kernel has already collapsed at the $k$ that variance
would pick.  What is missing is an \emph{a priori} budget: a geometric
estimate, computed classically on the data, of how many
coordinates the feature map can still bear.

This paper supplies that budget from fractal geometry and asks a
comparison that stays entirely on the quantum side: for the same
family of angle maps, the same fidelity kernel, the same algorithms
that read $K$, is it better to encode the attributes selected by
fractal dimension than to encode every attribute, the leading
principal components, or $k$ random axes?

\paragraph{Contributions.}
\begin{enumerate}
\item We state the tabular QML problem as a mismatch between embedding
  dimension and the Hilbert-space width of an angle map, and we
  separate that mismatch from the orthogonal question of sample size.
\item We define three diagnostics on $K$ (\emph{near}, \emph{far},
  mean off-diagonal) and an operational \emph{alive} rule, so that
  collapse is not confused with a lucky or unlucky SVM score at
  $n=32$.  The last width at which the rule still holds is the
  \emph{knee} of the $q$-sweep.
\item On nine data sets we compare $\lceil D_2\rceil$ with that knee,
  against the TwoNN estimator of \citet{facco2017estimating} and
  PCA-$95\%$ as rival ceilings, and we report the FD-ASE kernel at its
  selected width---not only the PCA view.
\item We test whether the ceiling is an artefact of one $ZZ$ layer or of
  the default angle scale: product-state $Z$ maps, a second $ZZ$ layer,
  a linear IQP map \citep{shepherd2009temporally}, packed dense-angle /
  re-uploading encodings \citep{perezsalinas2020data}, and a bandwidth
  sweep $c\in\{0.25,0.5,1,2\}$ in the sense of
  \citet{shaydulin2022importance,canatar2023bandwidth}.
\item A synthetic family with known width $k=1,\ldots,8$ checks that
  $D_2$ recovers $k$; wine and diabetes are re-run at $n=128$; the
  alive floors are varied by $\pm 50\%$; and the knee is repeated over
  ten kernel samples.
\item On IBM Quantum we reconstruct the one-layer $ZZ$ kernel at the
  fractal width and beyond it, and at $n=16$ we show that FD-ASE is
  the only selector that preserves the near/far geometry on the device.
\end{enumerate}

\section{Background}
\label{sec:background}

\subsection{Intrinsic dimension and fractal geometry}

The key insight of this paper is that the number of qubits should match
the \emph{intrinsic} dimension of the data, not its recorded width.
Intuitively, the intrinsic dimension counts how many independent
directions a data set actually uses.  A helix in
$\mathbb{R}^{3}$ is a one-dimensional object: every small
neighbourhood looks like a segment.  A sheet folded inside
$\mathbb{R}^{100}$ has intrinsic dimension two, even though the
recording instruments wrote one hundred columns.  The remaining
ninety-eight coordinates are redundant---they encode noise, correlations,
or measurement padding.

Classical machine learning has long exploited this gap.
\citet{camastra2002estimating} showed that estimating intrinsic
dimension before learning improves classifier design;
nearest-neighbour rules, manifold learners, and kernel methods all
degrade when the recorded width far exceeds the intrinsic one, a
phenomenon at the core of the curse of dimensionality.

\begin{figure}[t]
\centering
\includegraphics[width=0.92\textwidth]{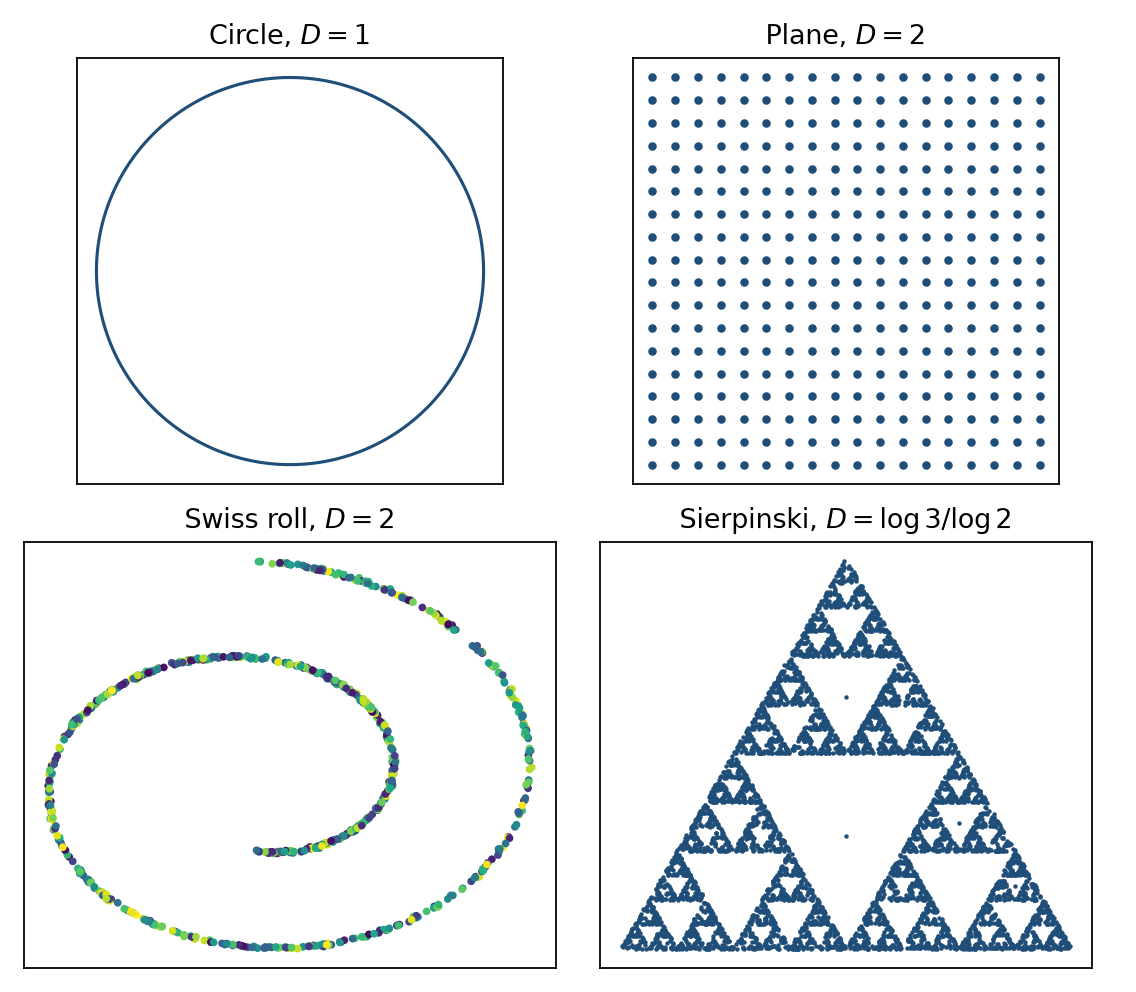}
\caption{Four sets whose dimension is not the number of recorded
coordinates.  A circle is a one-dimensional curve.  A plane is a
two-dimensional region.  A Swiss roll is a two-dimensional sheet that
occupies three linear directions.  The Sierpinski triangle repeats
itself at half the scale and has dimension $\log 3/\log 2\approx
1.58$.}
\label{fig:idexamples}
\end{figure}

Figure~\ref{fig:idexamples} separates the recorded width $E$ from the
intrinsic dimension.  A circle drawn as $(\cos t,\sin t)$ and padded
with eighteen noise coordinates has $E=20$ and intrinsic dimension
about $1$: one parameter $t$ traces the signal, and two of the twenty
coordinates carry it.  A plane in $100$ dimensions has dimension $2$,
and PCA agrees because the support is a subspace.  A Swiss roll also
has dimension $2$, but the variance spreads over three linear
directions, so the dimension of the sheet is not the variance of the
embedding.  PCA-95\% is the smallest number of principal components
that together explain 95\% of the variance; each component mixes every
recorded column.  Standardising a circle padded with Gaussian noise
can make that cut demand almost every axis, because each noise
direction then has unit variance.  Variance is not the dimension of
the support.

The intrinsic dimension is not the number of classes and, except in
the linear case, is not the rank of the covariance.  It need not be an
integer.  A set is fractal when a small piece has the same shape as
the whole, so the dimension can fall between two integers.  The
Sierpinski triangle in Figure~\ref{fig:idexamples} is three copies of
itself at half the scale, and its dimension is $\log 3/\log 2\approx
1.58$.

\subsection{\texorpdfstring{The correlation fractal dimension $D_2$}%
  {The correlation fractal dimension D2}}

Among the many estimators of intrinsic dimension, the
\emph{correlation fractal dimension} $D_2$ is especially well suited
to tabular data.  Introduced by
\citet{grassberger1983characterization} to characterise strange
attractors, $D_2$ measures how the number of point pairs within
distance $r$ scales as $r$ shrinks: $S(r)\propto r^{D_2}$.
\citet{mandelbrot1982fractal} established the broader framework of
fractal geometry; \citet{faloutsos1994fractal} brought it into
databases, showing that spatial data obey power laws whose exponent is
$D_2$; and \citet{belussi1995estimating} used $D_2$ to estimate query
selectivity in spatial indexes with the LiBOC box-count algorithm.
Always $D_2\le E$, with equality only if the data fills the unit
cube.  A redundant axis barely raises $D_2$: that invariance is the
test that feature selection exploits.

The fit uses several scales $r$, so $D_2$ is slower to treat local
embedding noise as extra dimension than TwoNN.
\citet{facco2017estimating} introduced TwoNN as a minimal-neighbourhood
estimator that needs only the two nearest neighbours; it is fast, but at
neighbour scale it can inflate the ID if the embedding noise is of that
order.  We estimate $D_2$ classically, once, on up to $4000$
points---not on the $n=32$ subsample that enters the quantum kernel.
Each column is scaled to the unit cube before the box-count.  The fit
uses eight dyadic scales $r=2^{-j}$ and the slope of $\log S(r)$ versus
$\log r$, discarding levels at which $S$ has already dropped to the
occupancy floor $1.5N$ (every point in its own cell).

\subsection{FD-ASE: feature selection by fractal dimension}
\label{sec:selector}

In data mining, $D_2$ has been used for feature selection, anomaly
detection, and indexing cost estimation.  \citet{traina2000fast}
proposed a fast feature selection method that uses the correlation
fractal dimension---computed via a multi-resolution box-counting
grid---to identify which attributes carry geometric information.
That method adds, at each step, the original column that most
increases the partial dimension.  We stop when the gain falls below
$\varepsilon=0.15$, so the subset is short when a few recorded
columns already carry $D_2$.  \citet{sousa2002fractal} extended the
same partial dimension into FD-ASE, which walks attributes in the
order they are stored and groups those that do not raise the partial
dimension.  That column-order walk is implemented in the released
code and is not used in any table here.  The reported subsets are
the forward-greedy pass, fit on the same geometry sample as $D_2$,
then encoded on the kernel sample.

The output is a set of \emph{original} columns, not linear mixtures.
That matters at encoding time: a qubit assigned to a selected coordinate
rotates by a measured attribute; a qubit assigned to a principal
component rotates by a blend of every attribute, padding included.
$k$ for PCA is a hyperparameter (95\% variance, or a round number of
qubits); $\lceil D_2\rceil$ is read off the geometry.

\subsection{Quantum feature maps and the fidelity kernel}

All maps in this paper prepare a state $\lvert\psi(x)\rangle$ from a
real vector $x$ by rotations, optionally followed by entangling gates.
Each coordinate is scaled independently to $[0, c\pi]$ and then fed to
the map.  Unless a section names $c$, we use the usual $ZZ$ scale
$c=1$.  That min-max step equalises every included axis: a
low-variance principal component is stretched to the same angular
range as a signal component, so padding that has been selected into
the map is as expensive in Hilbert space as the signal.
\citet{shaydulin2022importance} identified $c$ as the kernel
bandwidth and showed that shrinking it retards concentration;
\citet{canatar2023bandwidth} showed that the same hyperparameter
controls generalisation.  The fractal ceiling in this paper is
therefore a budget for a named map \emph{at a named $c$}, not a
universal integer.  The kernel is always the
fidelity~\eqref{eq:kernel} of those states; we never train a
variational classifier on the device.

\paragraph{$Z$ map (product states).}
Each qubit $i$ is rotated by a function of the single coordinate $x_i$;
there is no entangling gate.  The prepared state is a product.  Inner
products therefore factor across qubits, and orthogonality can
accumulate only as a product of single-qubit overlaps.  We expect this
map to concentrate more slowly as $q$ grows than an entangled map of the
same width.

\paragraph{$ZZ$ map.}
The construction of \citet{havlicek2019supervised} prepends Hadamards
and then applies single-qubit $Z$ rotations together with two-qubit
$ZZ$ interactions that depend on products of coordinates.  One layer
with linear (nearest-neighbour) entanglement is the default in the
ceiling sweep and on hardware.  A second layer repeats the diagonal
block: more entanglement, a more expressive map, and---we will see---a
kernel that is already geometrically dead at the fractal width.

\paragraph{IQP map.}
Instantaneous quantum polynomial circuits, introduced by
\citet{shepherd2009temporally} and used as feature maps by
\citet{havlicek2019supervised}, apply Hadamards, then a diagonal of
single-qubit phases $RZ(x_i)$ and controlled phases on products
$x_i x_j$.  We use a linear skeleton of those controlled phases
(adjacent pairs only), so the two-qubit cost stays comparable to linear
$ZZ$.  IQP is a second entangled family, not a packed encoding: still
one coordinate per qubit.

\paragraph{Packed maps.}
Dense-angle encoding writes two angles ($RY$, $RZ$) on each qubit per
layer.  Data re-uploading, proposed by \citet{perezsalinas2020data},
repeats a single-angle block so that extra coordinates share the same
wires.  These maps test a different objection: perhaps
$\lceil D_2\rceil$ is only a one-feature-per-qubit counting rule, and
PCA-95\% could still be encoded if several components share a qubit.
We keep $q$ fixed at the fractal ceiling and vary how many principal
components are stacked onto those wires.

On a simulator the kernel is the exact statevector fidelity.  On
hardware we estimate the same number by compute--uncompute: prepare
$U(x)\lvert 0\rangle$, apply $U(y)^\dagger$, and read the probability
of the all-zero bitstring.  No classifier runs on the QPU; the device
is used only to reconstruct $K$.

\subsection{Kernel diagnostics: near, far, alive, and the knee}
\label{sec:alive}

Accuracy of a precomputed SVM at $n=32$ is a noisy score.  It can jump
when a support vector moves, and it can look healthy on a kernel that is
already almost diagonal, or poor on a kernel that still has geometry.
We therefore judge the quantum object---$K$ itself---before we look at
downstream tasks.

\paragraph{Three numbers on one kernel.}
Write $d(x_i,x_j)$ for Euclidean distance in the \emph{encoded}
coordinates (the $q$ principal components, FD-ASE columns, or whichever
view the map sees), not in Hilbert space.  Among the $\binom{n}{2}$
unordered pairs, let $\mathcal{N}$ be the $20$ pairs of smallest $d$
and $\mathcal{F}$ the $20$ pairs of largest $d$ (or fewer if $n$ is
tiny).  Then
\begin{align}
  \mathrm{near}
  &= \frac{1}{\lvert\mathcal{N}\rvert}
     \sum_{(i,j)\in\mathcal{N}} K_{ij},
  \\
  \mathrm{far}
  &= \frac{1}{\lvert\mathcal{F}\rvert}
     \sum_{(i,j)\in\mathcal{F}} K_{ij}.
\end{align}
\emph{Near} asks: when two records are neighbours in the coordinates the
circuit actually sees, are their quantum states still similar?
\emph{Far} asks the opposite: when two records are far apart in those
coordinates, do the states still overlap?  If the feature map respects
the geometry of $X$, neighbours should remain similar as states and
distant points should not, so $\mathrm{near}$ should stand clearly
above $\mathrm{far}$.

The third diagnostic does not look at neighbours.  The mean off-diagonal
\begin{equation}
  \bar K \;=\; \frac{1}{n(n-1)}\sum_{i\neq j} K_{ij}
\end{equation}
is the average similarity of every distinct pair.  When the prepared
states become pairwise orthogonal, $\bar K\to 0$ and $K$ is the
identity for all practical purposes: every point looks equally unlike
every other point, and a kernel method has nothing left to work with.

\paragraph{When is the kernel alive?}
A kernel is declared \emph{alive} when three inequalities hold at once:
\begin{equation}
  \label{eq:alive}
  \mathrm{near}\ge 0.25,
  \qquad
  \frac{\mathrm{near}}{\mathrm{far}}\ge 2,
  \qquad
  \bar K \ge 0.03.
\end{equation}
Each clause blocks a different failure mode.  A large ratio with
$\mathrm{near}\approx 0$ is a ratio of two numerical zeros: the map has
already concentrated, and dividing two tiny fidelities is not geometry.
A large $\bar K$ with $\mathrm{near}\approx\mathrm{far}$ is a kernel
that still has mass off the diagonal but does not rank neighbours above
strangers (two-layer $ZZ$ at the fractal width fails this way;
Section~\ref{sec:ablation}).  A large $\mathrm{near}$ with
$\bar K\approx 0$ cannot occur for long: if neighbours are similar,
some off-diagonals are large.  The numerical floors are operational,
not a hypothesis test; they are held fixed for every data set, view,
and map in this paper.

\paragraph{What we call the knee.}
Plot $q$ on the horizontal axis and any of the geometric
diagnostics---or the binary alive bit---on the vertical axis.  For a
usable angle map the curve starts live at small $q$ and, as more
coordinates (and therefore more qubits) enter the state, the prepared
states become orthogonal and the diagnostics fall.  The \emph{knee} is
that break: the largest $q$ at which the kernel is still alive,
after which it is dead.  We also write \emph{last alive} $q$
for the same integer.  It is the number we compare with
$\lceil D_2\rceil$.

The three floors, $0.25$, $2$ and $0.03$, are fixed, but they are still a choice.
As a reading that does not use them, we also track the effective rank
of $K$ \citep{roy2007effective}, the exponential of the Shannon entropy
of its normalized eigenvalues.  A fidelity kernel near the identity has
a flat spectrum and effective rank near $n$.  A kernel that still
carries a few directions has a smaller effective rank.
Section~\ref{sec:erank} compares the two readings on breast and moons.

This is not the knee of a training-loss curve, nor the elbow of a
clustering scree plot.  Nothing is being optimised against $q$.  The
sweep only asks how wide a feature map can be before $K$ stops seeing
the shape of the data.

Table~\ref{tab:breast-metrics} is a worked reading on breast cancer
(one-layer $ZZ$, PCA view, $n=32$).  At $q=2$ the mean of $K$ is
still high ($0.30$), but near $0.49$ and far $0.34$ give a ratio
below $2$: neighbours and strangers look too alike, so the kernel is
\emph{not} alive.  At $q=3=\lceil D_2\rceil$ all three clauses pass
(near $0.27$, far $0.12$, $\bar K=0.14$).  At $q=4$ they all fail:
near has dropped to $0.11$, the ratio is again below $2$, and
$\bar K=0.07$ is sliding toward a diagonal kernel.  By $q=8$ nothing
remains off the diagonal.  The knee is $q=3$.  PCA-95\% would encode
$q=10$, well past the break.

\begin{table}[t]
\centering
\caption{How to read near, far and alive on one sweep.
Breast cancer, one-layer $ZZ$, PCA view, $n=32$.
$E=30$, $D_2\approx 2.49$, $\lceil D_2\rceil=3$, PCA-95\% would
encode $10$.
The kernel is alive only at $q=3$; that width is the knee of this
curve.}
\label{tab:breast-metrics}
\small
\begin{tabular}{@{}lcccc@{}}
\toprule
$q$ & near & far & mean $K$ & alive? \\
\midrule
$2$ & $0.49$ & $0.34$ & $0.30$ & no (ratio $<2$) \\
$\mathbf{3=\lceil D_2\rceil}$ & $0.27$ & $0.12$ & $0.14$
  & \textbf{yes} \\
$4$ & $0.11$ & $0.07$ & $0.07$ & no \\
$8$ & $0.00$ & $0.01$ & $0.00$ & no \\
\bottomrule
\end{tabular}
\end{table}

If a classical RBF kernel on the same $q$ coordinates stays alive while
the fidelity kernel dies, the collapse is of the quantum feature map,
not of the classical curse of dimensionality on those axes.
Conversely, downstream QML scores (fidelity $k$NN, precomputed SVM F1,
kernel--label alignment) are reported as \emph{consequences} of a live
or dead $K$, not as the definition of the knee.

\section{Related work}
\label{sec:related}

\paragraph{Quantum kernels and concentration.}
\citet{schuld2019quantum} showed that feature-map circuits define an
implicit kernel, and \citet{schuld2021supervised} established that
supervised quantum models are, in general, kernel methods.  The $ZZ$
map used here is the one defined in Section~\ref{sec:background}.
\citet{thanasilp2024exponential} proved that the fidelity kernel of
global feature maps concentrates exponentially in the number of qubits.
\citet{kairon2026equivalence} showed that the same concentration is
equivalent to a barren plateau of the variational circuit from which
the fidelity kernel is built: a plateau-free ansatz yields a kernel
that does not concentrate exponentially.  Their numerical example still
chooses the qubit count by PCA and changes the circuit, not the
coordinates selected from the geometry of the table.
\citet{kubler2021inductive} related the inductive bias of quantum
kernels to the dimension of the encoding space.
\citet{huang2021power} showed that quantum advantage in kernel methods
depends on the data encoding, and
\citet{agliardi2025mitigating} proposed covariant kernels and subspace
restrictions to slow concentration.
\citet{lei2024neural} also cut the input before the map is built, with
a filter and a search over circuits.  The budget here is read from the
geometry and does not search the circuit.
\citet{abbas2021power} defined an effective dimension
for quantum models from the Fisher information: a property of the
\emph{model}, not of the support of $X$.
\citet{gilfuster2024understanding} argued that uniform generalisation
bounds are the wrong language for current QML.
\citet{schnabel2025kernels} compared fidelity and projected kernels
across many encodings and found that the choice of encoding changes
the scores.  Projected kernels change how $K$ is evaluated, not which
coordinates enter the map.  None of these works use an
intrinsic-dimension estimator of the data as an a priori qubit budget.

\paragraph{Dimensionality reduction for QML.}
\citet{belis2025learning} learned reduced representations for quantum
classifiers via autoencoders; their qubit count is still chosen by
trial or by PCA variance thresholds.
\citet{hur2024neural} train the embedding circuit itself.
In both cases the width is a result of training, not a number read
from the data before any circuit is built.

\section{Proposed method}
\label{sec:method}

We propose a two-step \emph{a priori} qubit budget for angle-encoded
quantum kernels on tabular data:

\begin{enumerate}
\item \textbf{How many qubits.}  Compute $D_2$ on the classical
  data (up to $4000$ rows, eight resolution levels, unit-cube columns).
  The qubit ceiling is $q^{*}=\lceil D_2\rceil$.
\item \textbf{Which coordinates.}  On the same data, grow a subset by
  the original column that most increases the partial dimension, and
  stop when the gain falls below $\varepsilon=0.15$.  If that subset has at least two attributes,
  encode those columns.  Otherwise fall back to the first $q^{*}$
  principal components.  In the released code this is the default
  forward-greedy pass of the FD-ASE class, not the column-order pass.
\end{enumerate}

The comparison that tests this budget stays entirely on the quantum
side:

\begin{quote}
For the same QML model---the same family of angle maps, the same
fidelity kernel, the same algorithms that read $K$---is it better to
encode the attributes selected by fractal dimension than to encode
every attribute, the leading principal components, or $k$ random axes?
\end{quote}

On the feature maps we actually run, the fractal width is where the
quantum kernel remains usable, and the PCA-$95\%$ width is often where
it is already dead.

\section{Experimental protocol}
\label{sec:protocol}

Table~\ref{tab:stack} summarises the stack.  Two reductions must not be
confused.  $\lceil D_2\rceil$ cuts \emph{qubits}: how many coordinates
enter the map.  The small $n$ cuts \emph{rows}, on the simulator and on
the QPU: how many points enter the kernel.  The scientific claim is the
qubit cut.  We do not claim that fewer training points work better.

\begin{table}[t]
\centering
\caption{What this paper runs.  ``Better'' means the quantum kernel
stays geometrically usable, not that QSVM beats SVM.}
\label{tab:stack}
\small
\begin{tabular}{@{}p{0.28\textwidth}p{0.64\textwidth}@{}}
\toprule
Layer & This paper \\
\midrule
Feature maps (simulator) &
  $ZZ$, one layer (ceiling sweep);
  $ZZ$, two layers;
  $Z$ (product states);
  linear IQP;
  packed dense-angle and data re-uploading \\
Kernel & $K_{ij}=\bigl|\langle\psi(x_i)\mid\psi(x_j)
  \rangle\bigr|^{2}$
  (statevector; on IBM: compute--uncompute, one-layer $ZZ$ only) \\
QML on $K$ (simulator, $n=32$) & precomputed SVM (QSVM-style),
  fidelity $k$NN, spectral clustering, one-class SVM, kernel ridge \\
Classical control & RBF kernel on the \emph{same} $q$ coordinates \\
Feature views & FD-ASE columns; first $q$ PCs (sweep and PCA-95\%);
  prefix / full table; $q$ random columns \\
Hardware & one-layer $ZZ$ kernel, $n=8$, $256$ shots; no classifier on
  the QPU; $Z$, IQP, two-layer $ZZ$, dense-angle and re-uploading
  \emph{not} run on device (Section~\ref{sec:ablation}) \\
\midrule
Not used & a trained VQC; amplitude encoding;
  QSVM vs SVM as the research question \\
\bottomrule
\end{tabular}
\end{table}

\subsection{Data sets and the two samples}

Nine data sets are used.  Intrinsic2 is a circle in twenty dimensions,
two coordinates carrying $(\cos t,\sin t)$ and eighteen of them noise,
generated in the code.  Moons are two interleaving curves from
\texttt{sklearn.datasets.make\_moons}, padded with eighteen noise
columns.  Iris, wine, and the Wisconsin diagnostic breast-cancer table
are the scikit-learn copies of the UCI sets.  Digits are the
scikit-learn $8\times 8$ optical digits, not MNIST.  Pendigits is the
UCI pen-based set, training and test files together.  Diabetes is the
ten-attribute baseline table of \citet{efron2004least} as shipped by
scikit-learn, not the eight-attribute UCI Pima set.  OneBig is a
reduced eight-dimensional version of the separated-blob control in
\citet{appel2007pkdd}: one large Gaussian cluster, four small ones,
and uniform noise, with the label collapsed to large versus the rest.
The kernel on OneBig stays alive past $\lceil D_2\rceil$ because the
clusters are easy to separate.  Addresses are in the data-availability
statement.

For each data set we draw two independent subsamples.  The
\emph{geometry sample} has up to $4000$ rows and is used only
classically: LiBOC box-count $D_2$ \citep{belussi1995estimating} with
eight resolution levels, TwoNN \citep{facco2017estimating}, and the
number of principal components that explain 95\% of the variance after
standardisation.  The \emph{kernel sample} has $n=32$ rows by default.
Here $n$ is the number of rows, drawn uniformly without replacement by
a second random generator.  The two draws are not forced to be
row-disjoint.  The kernel sample is the only matrix that enters a
quantum circuit.  A fidelity kernel on $n$ points is one circuit per
unordered pair, $\binom{n}{2}=n(n-1)/2$.  At $n=32$ that is $496$
circuits; the full breast-cancer table ($569$ rows) would be about
$1.6\times 10^{5}$.  On the device, $n=8$ is $28$ circuits and $n=16$
is $120$.  Splitting the two samples avoids estimating $D_2$ on the
same tiny set that we then kernelise.  Which rows land in the kernel
sample moves the knee by about one qubit; Section~\ref{sec:robustness}
reports the median over ten draws.  The $n=32$ diagnostic is held
fixed so that the nine data sets, the map ablation and the packed
encodings remain comparable; it is not a claim that 32 is the right
sample size.  Wine and diabetes, which concentrate before the ceiling
at $n=32$, are repeated at $n=128$ (Section~\ref{sec:robustness}).
Labels are used solely for the downstream scores and for a stratified
$70/30$ split of those kernel points; $D_2$ and the forward selection
are unsupervised.

\subsection{Views, widths, and maps}

At each integer $q\in\{2,\ldots,8\}$ (capped by $E$ and by $n-1$) we
build four classical views of the kernel sample: the first $q$ principal
components; $q$ random original columns; the prefix of $q$ recorded
columns; and, when the forward selection of Section~\ref{sec:selector} returns
at least two attributes on the geometry sample, that selected subset
(its width is fixed by the algorithm, not swept).  Tables label this
subset FD-ASE because that is the class name in the code.  Each view
is encoded with the
one-layer linear $ZZ$ map at $c=1$,
and---on the same coordinates---with a Gaussian RBF kernel as control.
Packed encodings (Section~\ref{sec:packed}) keep $q=\lceil D_2\rceil$
and vary the number of principal components stacked onto those qubits,
on breast, moons and pendigits.  The map ablation
(Section~\ref{sec:ablation}) keeps the PCA view and $n=32$, and replaces
the one-layer $ZZ$ map by $Z$, two-layer $ZZ$, or linear IQP, sweeping
$q=2$ to $8$ on breast, moons, iris and pendigits.  The bandwidth sweep
(Section~\ref{sec:bandwidth}) keeps that $ZZ$ map and varies $c$.
A synthetic family with known width $k$
(Section~\ref{sec:synthetic}) asks whether $D_2$ recovers $k$ and
whether the $ZZ$ knee follows $k$.  The same cube, rotated so that
the latent axes are no longer the recorded columns
(Section~\ref{sec:oblique}), asks whether that budget survives a
linear mixture.  For each map we report the knee of
that sweep (Section~\ref{sec:alive}).

\subsection{Hardware}

Hardware jobs use the IBM Quantum processor \texttt{ibm\_fez} (Open
plan, $256$ shots).  With $n=8$ there are $28$ unordered pairs and
therefore $28$ compute--uncompute circuits per job.  We reconstruct $K$
for a one-layer linear $ZZ$ map on: isotropic blobs at $q=5$ and $q=7$;
breast principal components at $q=3$ and $q=7$; FD-ASE columns on the
same eight breast rows ($q=4$); and three random breast columns on those
rows.  The eight breast rows are a subset of the simulator's kernel
sample, so the device is asked to implement the same map on the same
points, not a new draw.  Classifiers from Table~\ref{tab:stack} are not
executed on the QPU.  Wall-clock was $36$--$48\,\mathrm{s}$ per job with
an empty queue.  We send a map to the device only when the simulator
knee sits next to $\lceil D_2\rceil$ (last alive
$q\le\lceil D_2\rceil+1$, and the map does die before $q=8$).
Product-state $Z$, two-layer $ZZ$ and IQP fail that gate
(Section~\ref{sec:ablation}); they stay on the simulator.

\paragraph{Code and data.}
Code, statevector sweep tables, figures, and the IBM kernel matrices
are publicly available.\footnote{\url{https://github.com/anapaulaappel/qml-qubit-budget}}
That repository also contains the $D_2$ box-count and the forward
selection used here.  Sources of the nine tables are given in the
data-availability statement.

\section{Results}
\label{sec:results}

\subsection{Simulator: the ceiling}
\label{sec:ceiling}

\begin{figure}[t]
\centering
\includegraphics[width=\textwidth]{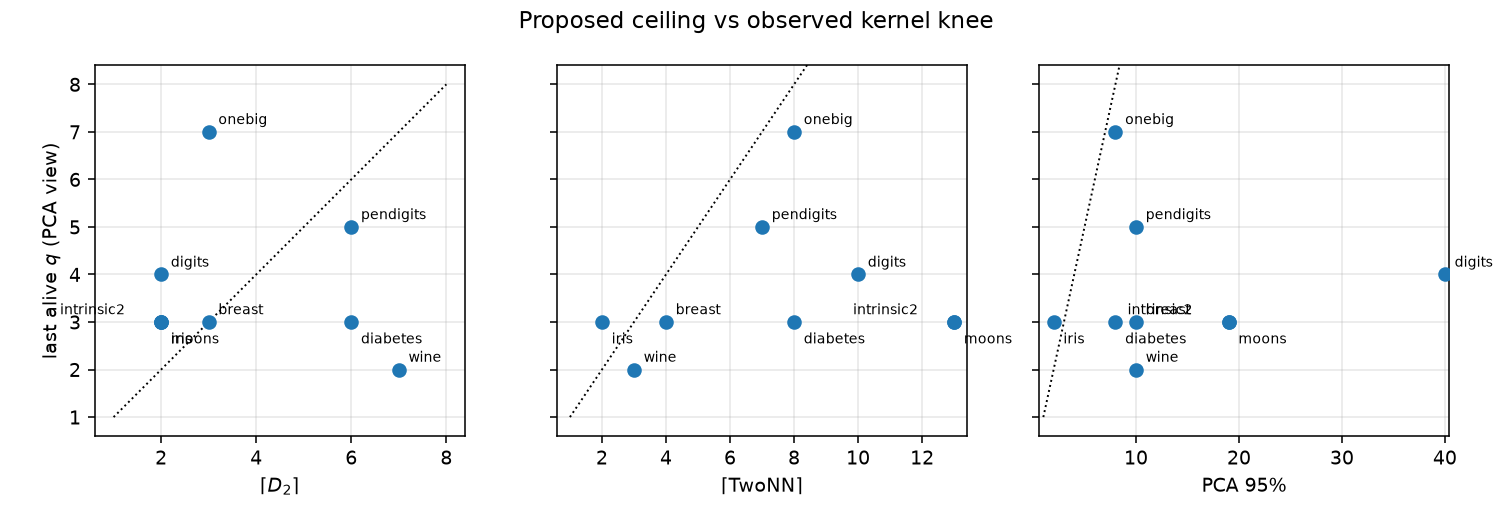}
\caption{A point on the dotted $y=x$ line means the proposed ceiling
equals the knee of the fidelity kernel (PCA view, one-layer $ZZ$,
$n=32$; knee $=$ last alive $q$, Section~\ref{sec:alive}).
\emph{Left:} $\lceil D_2\rceil$ sits next to that knee (breast on the
line at $3$; pendigits $6$ vs.\ $5$).
\emph{Centre:} TwoNN \citep{facco2017estimating} overshoots
moons/intrinsic2 ($\approx 13$) because it treats embedding noise as
extra dimension.
\emph{Right:} PCA-95\% asks for $10$--$40$ qubits while the kernel has
already died at $q\le 5$---that panel has its own $x$-axis so the live
knees ($2$--$7$) remain visible.  Outliers we report as failures: onebig
(kernel lives past the ceiling), wine/diabetes (high ID, $n=32$ dies
early), digits (unreliable box-count).}
\label{fig:ceilings}
\end{figure}

Table~\ref{tab:ceilings} reports, for each data set, $E$, $D_2$, the
proposed ceiling, the $q$ that PCA-95\% would encode, and the
knee---the last alive $q$ on the PCA view of the one-layer $ZZ$
fidelity kernel ($n=32$), in the sense of Section~\ref{sec:alive}.
Figure~\ref{fig:ceilings} plots those rival ceilings against that knee.
The $D_2$ panel shares scale with the knee; PCA-95\% is allowed its own
horizontal axis so that digits ($q=40$) does not squash breast ($q=3$).

\begin{table}[t]
\centering
\caption{Quantum-versus-quantum ceilings.  The last column is the knee
of the PCA view of the one-layer $ZZ$ fidelity kernel: the last $q$ at
which $K$ is still alive (Section~\ref{sec:alive}).}
\label{tab:ceilings}
\small
\begin{tabular}{@{}lrrrrr@{}}
\toprule
data set & $E$ & $D_2$ & $\lceil D_2\rceil$ & PCA-95\% & knee \\
\midrule
intrinsic2 & 20 & 0.72 & 2 & 19 & 3 \\
moons & 20 & 0.72 & 2 & 19 & 3 \\
iris & 4 & 1.92 & 2 & 2 & 3 \\
breast & 30 & 2.49 & 3 & 10 & \textbf{3} \\
diabetes & 10 & 5.82 & 6 & 8 & 3 \\
wine & 13 & 6.48 & 7 & 10 & 2 \\
digits & 64 & 0.91 & 2 & 40 & 4 \\
pendigits & 16 & 5.35 & 6 & 10 & 5 \\
onebig & 8 & 2.27 & 3 & 8 & 7 \\
\bottomrule
\end{tabular}
\end{table}

\begin{figure}[t]
\centering
\includegraphics[width=\textwidth]{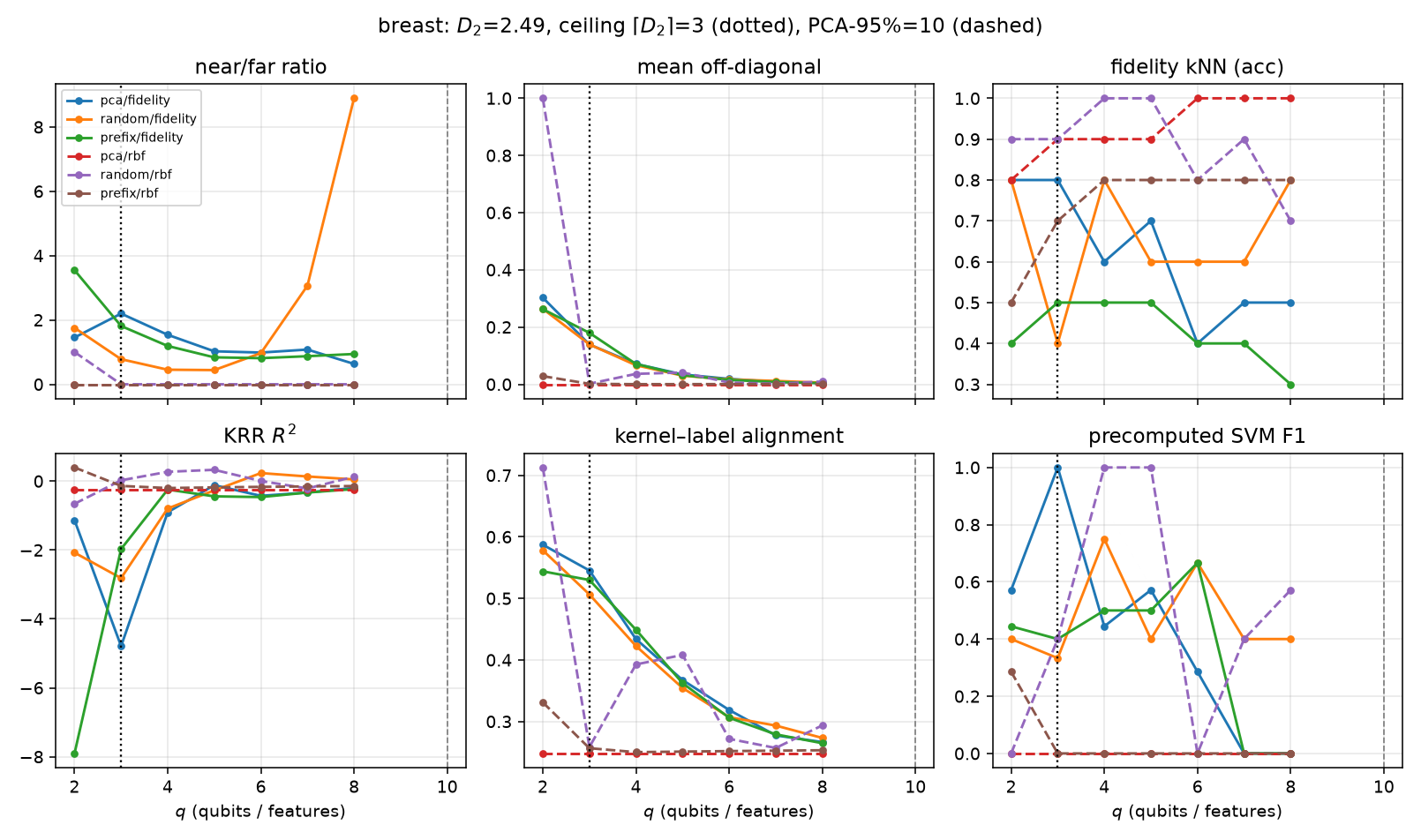}
\caption{Breast cancer ($E=30$, $D_2\approx 2.49$).  Dotted vertical:
fractal budget $q=3$.  Grey dashed: PCA-95\% ($q=10$).  Solid lines:
fidelity kernel (the QML object).  Dashed lines: classical RBF on the
same coordinates (control).  \emph{Mean off-diagonal} (top centre):
every fidelity curve falls toward $0$ as $q$ grows---states become
orthogonal; RBF does not, so this is the feature map, not classical
high dimension.  \emph{near/far}, \emph{$k$NN}, \emph{SVM F1}: on the
PCA view of the fidelity kernel the usable geometry and the QML scores
peak at the knee $q=3$ and degrade past it.  \emph{Alignment} falls
with $q$.  \emph{KRR} is noisy at $n=32$; we do not hang the claim on
it.  The practical reading: encoding $3$ PCs (the fractal width) keeps
a live quantum kernel and a working $k$NN/SVM on that kernel; encoding
$10$ PCs, as QML-practice PCA would, feeds those same algorithms a
kernel that is already diagonal.}
\label{fig:breast}
\end{figure}

Encoding as in QML-practice PCA kills the kernel where the fractal
keeps it alive (Figure~\ref{fig:ceilings}, right panel vs.\ left).
Breast (Figure~\ref{fig:breast}, Table~\ref{tab:breast-metrics}):
PCA-95\% asks for $10$ qubits; fidelity has already collapsed at $q=4$,
while at $q=3=\lceil D_2\rceil$---the knee---the same QSVM, $k$NN and
alignment still see geometry.  Moons / intrinsic2: PCA asks for $19$;
the angle map lives at $2$--$3$.  Pendigits sits next to the fractal
ceiling ($5$ vs.\ $6$), not at $10$ or $16$.  On breast the
random-column view of the fidelity kernel does not sustain the alive
rule in the $n=32$ sweep---random axes are not a substitute for FD-ASE
there.

Honest failures: OneBig (separated blobs, the kernel lives past
$\lceil D_2\rceil$); wine and diabetes (high ID, $n=32$ concentrates
\emph{before} the ceiling); digits ($E=64$, box-count on the floor---do
not trust a grid estimator when $E\gg\log N$).  TwoNN inflates moons /
intrinsic2 to $\approx 13$ because it sees the $18$ noise axes at
neighbour scale; $D_2$ does not.

\subsection{FD-ASE at its selected width}
\label{sec:fdase-table}

Table~\ref{tab:fdase} reports the one-layer $ZZ$ kernel on the columns
that FD-ASE selects from the geometry sample, encoded on the same
$n=32$ kernel sample as the PCA sweep.  This is the second step of the
proposed method, which Table~\ref{tab:ceilings} does not test.

On iris, diabetes and OneBig the selected subset is alive.  On breast
the algorithm returns four original columns---one above
$\lceil D_2\rceil$---and the kernel fails the ratio clause (near
$0.083$, far $0.074$).  That is consistent with the hardware $n=16$
job, where the same columns still rank neighbours above strangers
(ratio $2.1$) but do not pass the full alive rule.  On wine and
pendigits FD-ASE returns eight columns, past the $ZZ$ knee, and the
kernel is dead.  Intrinsic2 and moons have no FD-ASE subset: after
padding with independent noise the partial dimension never saturates
on a small original-column basis.  The honest reading is that
$\lceil D_2\rceil$ is the qubit budget; FD-ASE is a selector of
original columns that realises that budget when it returns a short
subset, not a guarantee that every selected subset is alive.

\begin{table}[t]
\centering
\caption{FD-ASE view, one-layer $ZZ$, $n=32$.  A dash means FD-ASE did
not return at least two columns.  PCA knee is from
Table~\ref{tab:ceilings}.}
\label{tab:fdase}
\small
\begin{tabular}{@{}lrrrrcc@{}}
\toprule
data & $\lceil D_2\rceil$ & PCA knee & FD-ASE $q$ & near & ratio & alive \\
\midrule
intrinsic2 & 2 & 3 & --- & --- & --- & --- \\
moons & 2 & 3 & --- & --- & --- & --- \\
iris & 2 & 3 & 3 & $0.43$ & $4.5$ & yes \\
breast & 3 & 3 & 4 & $0.08$ & $1.1$ & no \\
diabetes & 6 & 3 & 2 & $0.98$ & $3.1$ & yes \\
wine & 7 & 2 & 8 & $0.01$ & $1.8$ & no \\
digits & 2 & 4 & 2 & $0.80$ & $1.7$ & no \\
pendigits & 6 & 5 & 8 & $0.04$ & $22.8$ & no \\
onebig & 3 & 7 & 4 & $0.75$ & $11.0$ & yes \\
\bottomrule
\end{tabular}
\end{table}

\subsection{Bandwidth}
\label{sec:bandwidth}

\begin{figure}[t]
\centering
\includegraphics[width=\textwidth]{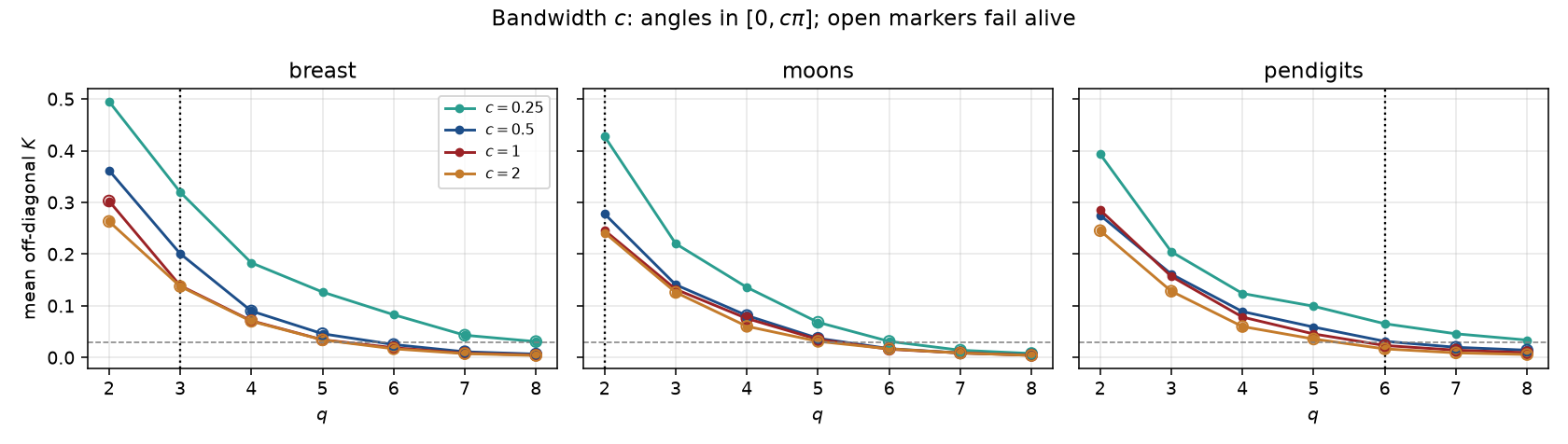}
\caption{Mean off-diagonal of $K$ versus $q$ at four bandwidths.
Dotted vertical: $\lceil D_2\rceil$.  Dashed horizontal: the $0.03$
alive floor.  Open markers fail the alive rule.  Smaller $c$ (angles
in $[0,c\pi]$) delays collapse; $c=1$ is the default $ZZ$ scale.}
\label{fig:bandwidth}
\end{figure}

The default map scales each included coordinate to $[0,\pi]$.
Figure~\ref{fig:bandwidth} and Table~\ref{tab:bandwidth} repeat the
PCA $q$-sweep at $c\in\{0.25,0.5,1,2\}$.  On breast, $c=1$ has knee
$3=\lceil D_2\rceil$; $c=0.5$ stays at $3$; $c=0.25$ lives through
$q=6$; $c=2$ has no knee in $q=2$--$8$.  Moons and pendigits show the
same monotone: shrinking $c$ moves the knee later, stretching it
kills the kernel earlier.  A critic who says ``PCA-95\% would live if
you tuned the bandwidth'' is right that $c$ can buy extra qubits, and
wrong that this removes the budget.  Tuning $c$ changes the map, so
the integer $\lceil D_2\rceil$ must be quoted together with $c$.  At
the $ZZ$ scale used in practice ($c=1$), the knee still sits next to
the fractal width on these three data sets.

\begin{table}[t]
\centering
\caption{Last alive $q$ versus bandwidth $c$ (PCA view, one-layer
$ZZ$, $n=32$).  A dash means the alive rule never held.}
\label{tab:bandwidth}
\small
\begin{tabular}{@{}lrrrrr@{}}
\toprule
data & $\lceil D_2\rceil$ & $c=0.25$ & $c=0.5$ & $c=1$ & $c=2$ \\
\midrule
breast & 3 & 6 & 3 & \textbf{3} & --- \\
moons & 2 & 4 & 3 & 3 & 2 \\
pendigits & 6 & 8 & 6 & 5 & --- \\
\bottomrule
\end{tabular}
\end{table}

\subsection{Synthetic width $k$}
\label{sec:synthetic}

To isolate intrinsic width from the idiosyncrasies of UCI tables we
generate $k$ independent coordinates in $[-1,1]$, embed them in
$E=20$ with null padding, and estimate $D_2$ on $n=400$ rows.
Table~\ref{tab:synthetic} shows that box-count $D_2$ recovers $k$
($0.95$ at $k=1$ through $7.28$ at $k=8$) and that PCA-95\% equals
$k$.  The one-layer $ZZ$ knee on the prefix view, at $n=32$ and
$c=1$, does \emph{not} follow $k$: once $k\ge 3$ the kernel dies at
$q=2$--$3$, the same window as wine and diabetes.  Two independent
full-range axes remain alive even with extra constant qubits (knee
$8$ at $k=2$).  The estimator sees the geometry; the map cannot
encode more than a handful of full-range angles at this $n$ and $c$.
That is the scope condition already visible on high-ID tables, now
with a known $k$.

\begin{table}[t]
\centering
\caption{Synthetic $k$-width in $E=20$.  Knee: last alive $q$ of
one-layer $ZZ$ on the prefix view, $n=32$, $c=1$.}
\label{tab:synthetic}
\small
\begin{tabular}{@{}rrrrr@{}}
\toprule
$k$ & $D_2$ & $\lceil D_2\rceil$ & PCA-95\% & knee \\
\midrule
$1$ & $0.95$ & $2$ & $1$ & --- \\
$2$ & $1.83$ & $2$ & $2$ & $8$ \\
$3$ & $2.63$ & $3$ & $3$ & $3$ \\
$4$ & $3.65$ & $4$ & $4$ & $2$ \\
$5$ & $4.89$ & $5$ & $5$ & $2$ \\
$6$ & $5.79$ & $6$ & $6$ & $3$ \\
$7$ & $6.56$ & $7$ & $7$ & $3$ \\
$8$ & $7.28$ & $8$ & $8$ & $2$ \\
\bottomrule
\end{tabular}
\end{table}

\subsection{Oblique axes}
\label{sec:oblique}

\begin{figure}[t]
\centering
\includegraphics[width=0.72\textwidth]{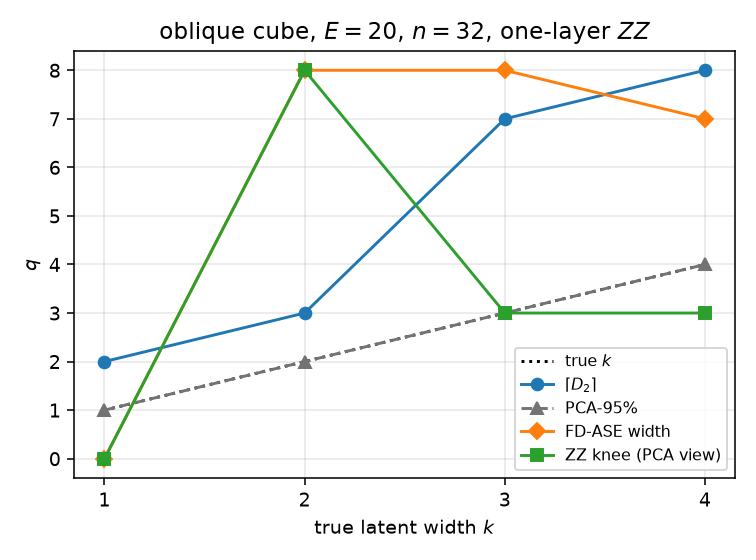}
\caption{Rotated $k$-cube.  $\lceil D_2\rceil$ and the FD-ASE width
leave the true $k$; PCA-95\% stays on it.  The $ZZ$ knee is read on
the PCA view.}
\label{fig:oblique}
\end{figure}

The cube in Section~\ref{sec:synthetic} sits on the first $k$
coordinate axes.  Forward selection on the partial dimension is built
for that case: a redundant recorded
column barely raises the partial dimension.  A linear mixture is the
other case.  We take the same null-padded cube and rotate it by one
Haar orthogonal matrix, so the support is still $k$-dimensional but
no recorded column is a latent axis.  Table~\ref{tab:oblique} and
Figure~\ref{fig:oblique} use the same one-layer $ZZ$ kernel, $n=32$,
$c=1$.

Box-count $D_2$ no longer recovers $k$ once $k\ge 2$.  The estimator
scales each column onto the unit cube before counting boxes; after a
rotation that scaling shears the set, and the slope rises ($2.64$ at
$k=2$, $6.33$ at $k=3$, $7.59$ at $k=4$, against $1.83$, $2.63$ and
$3.65$ on the aligned cube).  PCA-$95\%$ is invariant to the rotation
and equals $k$ on every row.  FD-ASE, reading the sheared columns,
returns seven or eight attributes for $k=2,3,4$, and the fidelity
kernel on that subset is dead.  At $k=1$ it returns fewer than two
columns.  The PCA view of the same kernel behaves like the aligned
sweep: at $k=2$ the knee is still $8$, and the kernel is alive at the
inflated ceiling $q=3$; at $k=3$ and $k=4$ the knee is $3$, so the
fractal ceilings $7$ and $8$ are already past the collapse.

When the support is a rotated linear subspace, the qubit count that
matches the geometry is the covariance rank, and the coordinates to
encode are principal components.  The aligned tables in the rest of
this paper are the case FD-ASE is built for.

\begin{table}[t]
\centering
\caption{Rotated $k$-cube in $E=20$.  Knee: last alive $q$ of the
one-layer $ZZ$ kernel on the PCA view, $n=32$, $c=1$.  A dash means
FD-ASE returned fewer than two columns, or that the alive rule never
held.}
\label{tab:oblique}
\small
\begin{tabular}{@{}rrrrrrr@{}}
\toprule
$k$ & $D_2$ & $\lceil D_2\rceil$ & PCA-95\% & FD-ASE & alive? & knee \\
\midrule
$1$ & $0.95$ & $2$ & $1$ & --- & --- & --- \\
$2$ & $2.64$ & $3$ & $2$ & $8$ & no & $8$ \\
$3$ & $6.33$ & $7$ & $3$ & $8$ & no & $3$ \\
$4$ & $7.59$ & $8$ & $4$ & $7$ & no & $3$ \\
\bottomrule
\end{tabular}
\end{table}

\subsection{Robustness: seeds, floors, and $n=128$}
\label{sec:robustness}

\begin{figure}[t]
\centering
\includegraphics[width=0.55\textwidth]{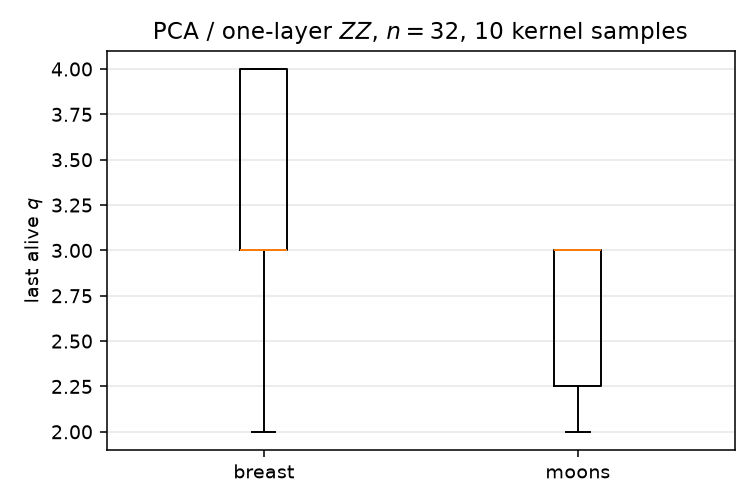}
\caption{Last alive $q$ over ten kernel samples ($n=32$, PCA,
one-layer $ZZ$, $c=1$).  Median $3$ on both breast and moons.}
\label{fig:seeds}
\end{figure}

The knee is an integer read from one kernel sample.  Ten independent
draws of the $n=32$ sample, PCA view, one-layer $ZZ$, $c=1$, give a
median last-alive $q$ of $3$ on both breast and moons
(Figure~\ref{fig:seeds}).  Breast ranges from $2$ to $4$ (two draws
have no knee); moons ranges from $2$ to $3$.  The default breast
result $q=3$ is the median, not a singleton.

Re-scoring the existing PCA/fidelity diagnostics with the three alive
floors multiplied by $0.5$ or $1.5$ leaves the breast, moons,
pendigits and intrinsic2 knees unchanged at the default $(0.25,2,0.03)$
and at a $50\%$ lower mean floor.  Raising the near floor to $0.38$ or
the ratio floor to $3$ can shift the knee by one qubit or, on breast,
declare no knee.  The rule is operational, as stated; it is not
invisible to its constants.

Wine and diabetes concentrate before $\lceil D_2\rceil$ at $n=32$.
Repeating the PCA sweep at $n=128$ moves the wine knee from $2$ to
$4$ and the diabetes knee from $3$ to $5$ (ceilings $7$ and $6$).
Larger $n$ helps; it does not take those two tables to the fractal
width.  The synthetic $k$-sweep (Section~\ref{sec:synthetic}) already
showed why: at $c=1$ the one-layer $ZZ$ map dies around three
full-range coordinates even when $D_2$ is large.

\subsection{Downstream scores at $n=128$}
\label{sec:downstream}

\begin{figure}[t]
\centering
\includegraphics[width=\textwidth]{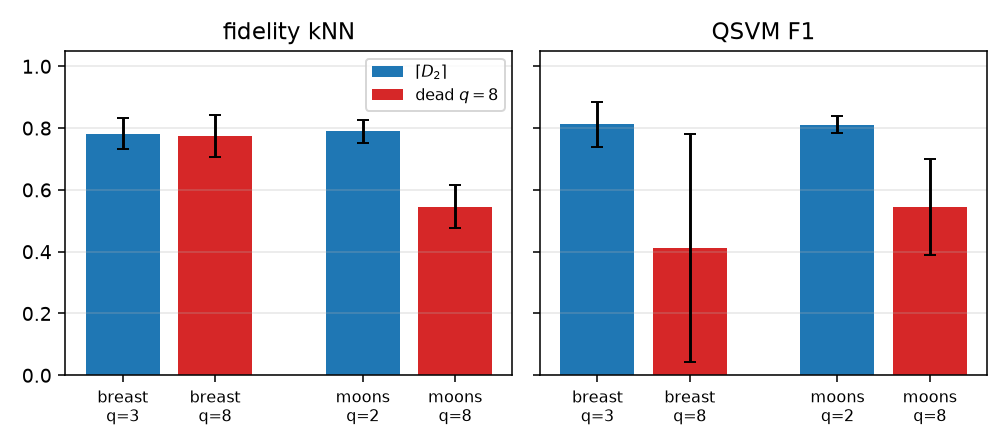}
\caption{Downstream scores at $n=128$, ten kernel samples.  Blue:
$\lceil D_2\rceil$ (alive on every draw).  Red: $q=8$ (dead on every
draw).  Error bars are one standard deviation.  Moons drops at the dead
width.  Breast $k$NN does not; breast SVM F1 drops on average and
varies widely.}
\label{fig:downstream}
\end{figure}

The $n=32$ breast curve is one test fold of about ten points.  Here the
kernel sample is $n=128$, with a stratified $70/30$ split, redrawn for
ten seeds.  The widths are the published ceilings---$q=3$ on breast and
$q=2$ on moons, from the seed-$0$ geometry sample---and $q=8$, the end
of the same PCA sweep, where that curve is already diagonal.
Table~\ref{tab:downstream} and Figure~\ref{fig:downstream} report
fidelity $k$NN accuracy and precomputed SVM F1.  Both are consequences
of $K$, not a new definition of the knee.

The alive bit is stable.  The ceiling width is alive on all ten draws;
$q=8$ is dead on all ten.  Mean off-diagonal fidelity is $0.30$ against
$0.013$ on breast, and $0.27$ against $0.004$ on moons.

On moons both tasks drop when the kernel dies: $k$NN from
$0.79\pm 0.04$ to $0.55\pm 0.07$, SVM F1 from $0.81\pm 0.03$ to
$0.55\pm 0.15$.  On breast they do not tell the same story.  Fidelity
$k$NN stays at $0.78\pm 0.05$ and $0.77\pm 0.07$.  SVM F1 falls on
average, from $0.81\pm 0.07$ to $0.41\pm 0.37$, but the dead draws
range from $0$ to $0.92$.  The single $n=32$ breast reading, F1 $=1$ at
$q=3$ and F1 $=0$ at $q=8$, was that small fold.  A dead kernel at
$n=128$ is a stable geometric fact.  It is not, on breast, a stable
drop in every downstream score.

\begin{table}[t]
\centering
\caption{Fidelity $k$NN accuracy and precomputed SVM F1 at $n=128$,
PCA view, one-layer $ZZ$, $c=1$.  Mean $\pm$ standard deviation over
ten kernel samples.  The ceiling is alive on every draw; $q=8$ is dead
on every draw.}
\label{tab:downstream}
\small
\begin{tabular}{@{}llccc@{}}
\toprule
data set & $q$ & alive & $k$NN & SVM F1 \\
\midrule
breast & $3=\lceil D_2\rceil$ & $10/10$ & $0.78\pm 0.05$ & $0.81\pm 0.07$ \\
breast & $8$ & $0/10$ & $0.77\pm 0.07$ & $0.41\pm 0.37$ \\
moons & $2=\lceil D_2\rceil$ & $10/10$ & $0.79\pm 0.04$ & $0.81\pm 0.03$ \\
moons & $8$ & $0/10$ & $0.55\pm 0.07$ & $0.55\pm 0.15$ \\
\bottomrule
\end{tabular}
\end{table}

\subsection{Effective rank, without the alive floors}
\label{sec:erank}

\begin{figure}[t]
\centering
\includegraphics[width=\textwidth]{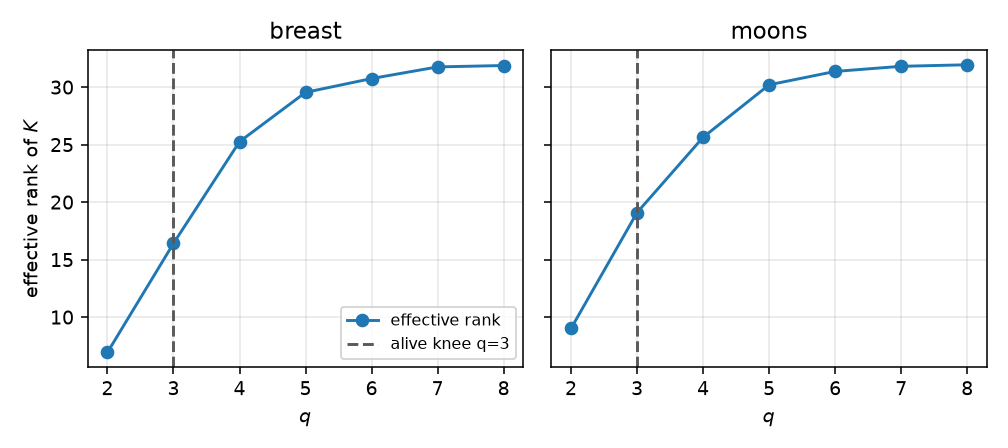}
\caption{Effective rank of $K$ against $q$.  The dashed line is the
alive knee ($q=3$).  The rank climbs toward $n=32$ with no corner
there.}
\label{fig:erank}
\end{figure}

On the same PCA view, one-layer $ZZ$ map, and $n=32$ draw used for the
ceiling, the effective rank of $K$ rises toward $n$ at every step
(Table~\ref{tab:erank}, Figure~\ref{fig:erank}).  There is no corner at
the alive knee.  The largest single step, on both breast and moons, is
from $q=2$ to $q=3$: $6.9\to 16.4$ and $9.0\to 19.1$.  That step is
where breast \emph{becomes} alive, not where it dies.  Death of the
alive rule is the next step, $q=3$ to $q=4$, where the effective rank
is already $25.3$ and $25.6$ out of $32$.  By $q=8$ both spectra are
flat ($31.9$).

The integer we compare with $\lceil D_2\rceil$ stays the alive knee.
The effective rank agrees with that knee only in direction: past
$q=3$ the kernel is spectrally close to the identity, which is the same
side of the collapse as the alive rule.  It does not replace the
three floors with another integer.

\begin{table}[t]
\centering
\caption{Effective rank of the one-layer $ZZ$ kernel, PCA view,
$n=32$.  The alive knee is $q=3$ on both sets.}
\label{tab:erank}
\small
\begin{tabular}{@{}rcccc@{}}
\toprule
$q$ & breast rank & alive? & moons rank & alive? \\
\midrule
$2$ & $6.9$ & no & $9.0$ & yes \\
$3$ & $16.4$ & yes & $19.1$ & yes \\
$4$ & $25.3$ & no & $25.6$ & no \\
$5$ & $29.5$ & no & $30.2$ & no \\
$6$ & $30.7$ & no & $31.4$ & no \\
$7$ & $31.8$ & no & $31.8$ & no \\
$8$ & $31.9$ & no & $31.9$ & no \\
\bottomrule
\end{tabular}
\end{table}

\subsection{Packed encodings: more than one attribute per qubit}
\label{sec:packed}

\begin{figure}[t]
\centering
\includegraphics[width=\textwidth]{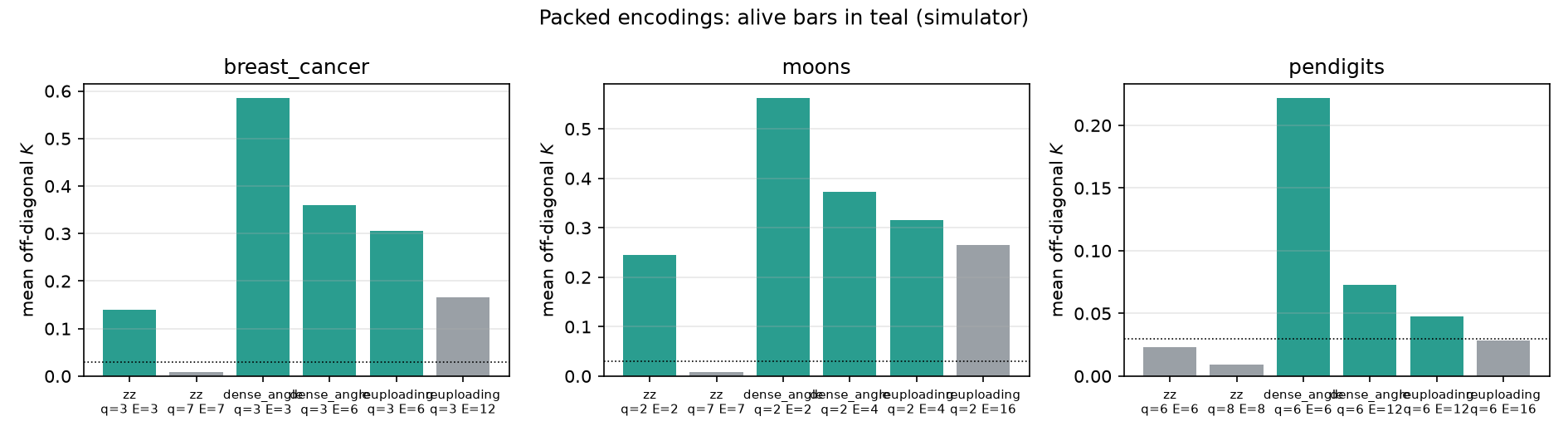}
\caption{Packed encodings, simulator, $n=32$.  $Y$: mean off-diagonal
of the fidelity kernel (higher $=$ less collapsed).  Teal: alive.
Grey: dead.  Dotted: $0.03$ threshold.  Left to right in each panel:
$ZZ$ at $\lceil D_2\rceil$; $ZZ$ beyond it; dense-angle / re-upload of
a few extra PCs on the fractal $q$; re-upload of a PCA-95\%-scale
feature count on that $q$.  The last of those is grey: stacking extra
coordinates onto fewer qubits is not a workaround.}
\label{fig:packed}
\end{figure}

The ceiling sweep used a one-layer $ZZ$ map (one coordinate per qubit).
Dense-angle and re-uploading \citep{perezsalinas2020data} were tested on
the statevector simulator only ($n=32$; breast, moons, pendigits).  The
objection they answer is whether one could still encode PCA-95\%, only
stacked onto $\lceil D_2\rceil$ wires.  Figure~\ref{fig:packed} and
Table~\ref{tab:packed} test that.

Each bar is the mean off-diagonal of $K$ ($n=32$).  Teal: the kernel
still passes the alive rule.  Grey: it does not.  The dotted line is
the $0.03$ floor.  On breast and moons, $ZZ$ at the fractal $q$ is
alive; $ZZ$ well beyond it is dead (first two bars).  Dense-angle and
re-uploading at that \emph{same} $q$, with two PCA coordinates per
qubit, stay teal---so $\lceil D_2\rceil$ is a qubit budget, not a
one-to-one accounting rule.  The last bar packs toward PCA-95\% onto
those qubits (12 PCs on 3 breast qubits; 16 PCs on 2 moons qubits) and
turns grey; on moons near/far drops below $1$ (geometry inverted:
distant points look more similar than neighbours).  Verdict: working at
the fractal width is better than encoding the PCA-95\% (or full) width;
one cannot recover that extra width by stacking it onto
$\lceil D_2\rceil$ qubits.

\begin{table}[t]
\centering
\caption{Packed encodings, $n=32$, PCA coordinates.}
\label{tab:packed}
\small
\begin{tabular}{@{}llrrrc@{}}
\toprule
data & encoding & $q$ & feats & mean $K$ & alive \\
\midrule
breast & $ZZ$ 1:1 & 3 & 3 & 0.139 & yes \\
breast & $ZZ$ 1:1 & 7 & 7 & 0.008 & no \\
breast & dense-angle & 3 & 6 & 0.360 & yes \\
breast & re-upload & 3 & 6 & 0.305 & yes \\
breast & re-upload & 3 & 12 & 0.165 & no \\
moons & $ZZ$ 1:1 & 2 & 2 & 0.246 & yes \\
moons & re-upload & 2 & 16 & 0.266 & no \\
pendigits & $ZZ$ 1:1 & 6 & 6 & 0.023 & no \\
pendigits & dense-angle & 6 & 12 & 0.073 & yes \\
\bottomrule
\end{tabular}
\end{table}

\subsection{Feature-map ablation: does the ceiling follow the map?}
\label{sec:ablation}

\begin{figure}[t]
\centering
\includegraphics[width=\textwidth]{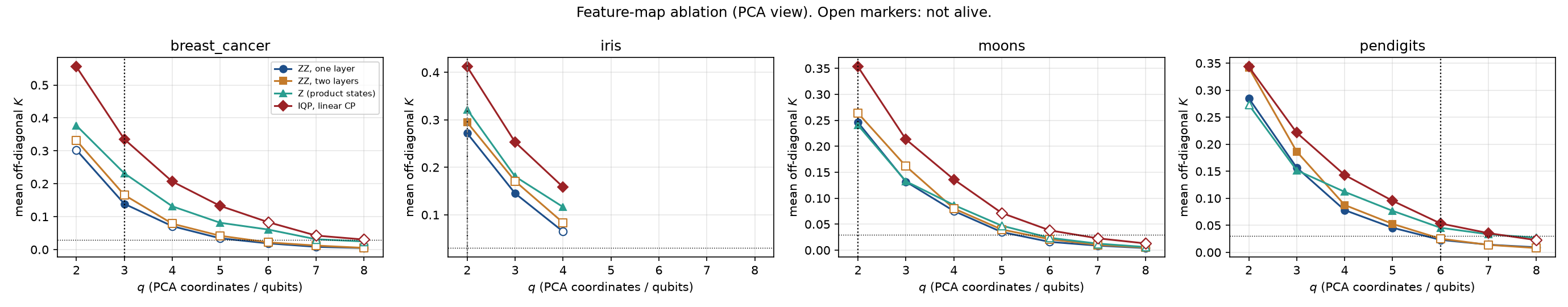}
\caption{Feature-map ablation, PCA view, $n=32$.  $Y$: mean
off-diagonal of $K$.  Dotted vertical: $\lceil D_2\rceil$.  Dashed
vertical: PCA-95\%.  Open markers fail the alive rule.  One-layer $ZZ$
dies at the fractal width on breast; two-layer $ZZ$ is already dead
there; $Z$ and IQP decay more slowly and overshoot
$\lceil D_2\rceil$.}
\label{fig:ablation}
\end{figure}

If $\lceil D_2\rceil$ were only a count of recorded columns, every angle
map of width $q$ should fail at the same $q$.  Figure~\ref{fig:ablation}
and Table~\ref{tab:ablation} keep the PCA view and $n=32$, and change
the map: one-layer $ZZ$ (the ceiling sweep), two-layer $ZZ$,
product-state $Z$, and linear IQP \citep{shepherd2009temporally}.  Open
markers are widths that fail the alive rule.  We run four data sets:
breast ($\lceil D_2\rceil=3$), moons ($2$), iris ($2$), and pendigits
($6$).

The one-layer $ZZ$ knee tracks the fractal ceiling on every data set:
breast dies after $q=3$, moons after $q=3$, iris after $q=3$,
pendigits after $q=5$ (one below $\lceil D_2\rceil=6$).  A second $ZZ$
layer either matches the one-layer knee (iris at $2$, pendigits at $5$)
or has no knee at all (breast and moons: the alive rule never holds in
$q=2$--$8$).  At breast $q=3$, two-layer $ZZ$ still has mean
off-diagonal $0.17$, but near $0.25$ and far $0.22$---neighbours are no
longer twice as similar as strangers.  Extra diagonal entanglement
flattens the geometry \emph{at} the fractal width; it does not buy a
wider live kernel.

The product-state $Z$ map is the opposite: on breast it remains alive
through $q=6$, on pendigits through $q=7$.  Without entanglement,
orthogonality accumulates only as a product of single-qubit overlaps, so
the Hilbert-space width can exceed $D_2$ for a while.  Linear IQP sits
in between (last alive $q=5$ on breast, $q=4$ on moons and iris, $q=7$
on pendigits).

The pattern is consistent across four data sets with
$\lceil D_2\rceil$ ranging from $2$ to $6$: the fractal number is a
budget for the map we actually intend to run, not a universal integer
that every circuit must hit.  For the entangled $ZZ$ layer used in QML
practice and on our IBM jobs, that budget is tight.  We therefore do
\emph{not} spend QPU time on $Z$, two-layer $ZZ$, or IQP: their
simulator knees do not sit next to $\lceil D_2\rceil$, which was the
gate for a hardware confirmation.

\begin{table}[t]
\centering
\caption{Last alive $q$ (the knee) on the PCA view, $n=32$.  A dash
means the alive rule never held in $q=2$--$8$, so that map has no knee
in the sweep.}
\label{tab:ablation}
\small
\begin{tabular}{@{}lrrrrr@{}}
\toprule
data & $\lceil D_2\rceil$ & $ZZ$ (1 layer) & $ZZ$ (2 layers)
  & $Z$ & IQP \\
\midrule
breast & 3 & \textbf{3} & --- & 6 & 5 \\
moons & 2 & 3 & --- & 4 & 4 \\
iris & 2 & 3 & 2 & 4 & 4 \\
pendigits & 6 & 5 & 5 & 7 & 7 \\
\bottomrule
\end{tabular}
\end{table}

\subsection{Hardware validation}
\label{sec:hardware}

Nine Sampler jobs on \texttt{ibm\_fez} (30 August 2026), one-layer $ZZ$
map, linear entanglement, $256$ shots.  Six jobs use $n=8$ ($28$
compute--uncompute circuits); three use $n=16$ ($120$ circuits) to
bring the sample size into the range where near/far can be scored.
With $256$ shots the sampling floor on a single probability is
$1/256\approx 0.004$; several MAEs in Table~\ref{tab:hw} sit on that
floor, so we do not interpret a $10^{-3}$ residual as device error
beyond shot noise.  No readout mitigation is applied.  The
classifiers of Table~\ref{tab:stack} are not executed on the QPU:
hardware only estimates $K$.

\begin{table}[t]
\centering
\caption{Hardware versus exact statevector.  Off-diag.\ $=$ mean of $K$
off the diagonal.}
\label{tab:hw}
\small
\begin{tabular}{@{}lrrrr@{}}
\toprule
job & $q$ & MAE & off-diag HW & off-diag SV \\
\midrule
blobs & 5 & 0.077 & 0.234 & 0.315 \\
blobs & 7 & 0.054 & 0.142 & 0.202 \\
breast PCA & 3 & 0.021 & 0.082 & 0.081 \\
breast PCA & 7 & 0.003 & 0.004 & 0.003 \\
breast FD-ASE & 4 & 0.012 & 0.065 & 0.068 \\
breast random & 3 & 0.030 & 0.164 & 0.166 \\
\bottomrule
\end{tabular}
\end{table}

\begin{figure}[t]
\centering
\includegraphics[width=\textwidth]{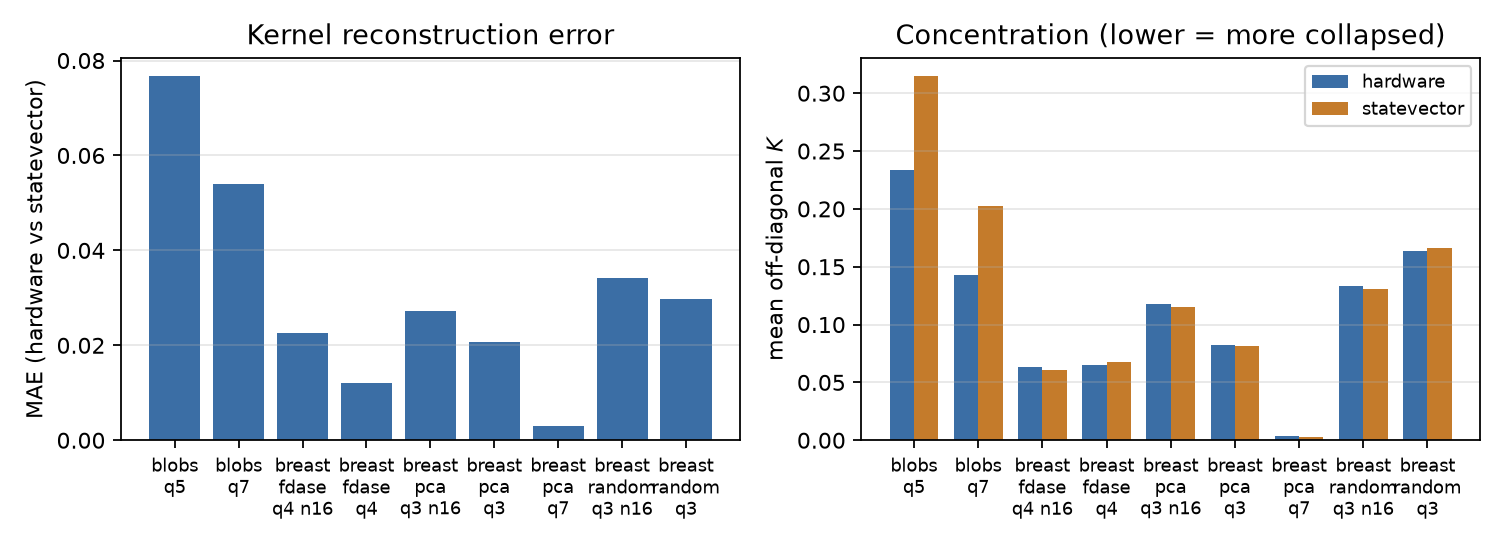}
\caption{IBM \texttt{ibm\_fez}, $256$ shots, $n=8$.  \emph{Left:} MAE
of the measured kernel versus the exact statevector---how faithfully the
device implements the map.  \emph{Right:} mean off-diagonal on hardware
(blue) vs.\ simulator (orange).  Breast PCA at $q=3$ and FD-ASE at
$q=4$ have low MAE and a non-zero kernel; at $q=7$ both bars sit on the
floor: the collapse is already in the exact $K$, the QPU is not adding
it.  Random $q=3$ has higher mean $K$ (not collapsed) but $n=8$ is too
small to score near/far; that comparison belongs to the $n=32$
simulator.}
\label{fig:hwbar}
\end{figure}

\begin{figure}[t]
\centering
\includegraphics[width=\textwidth]{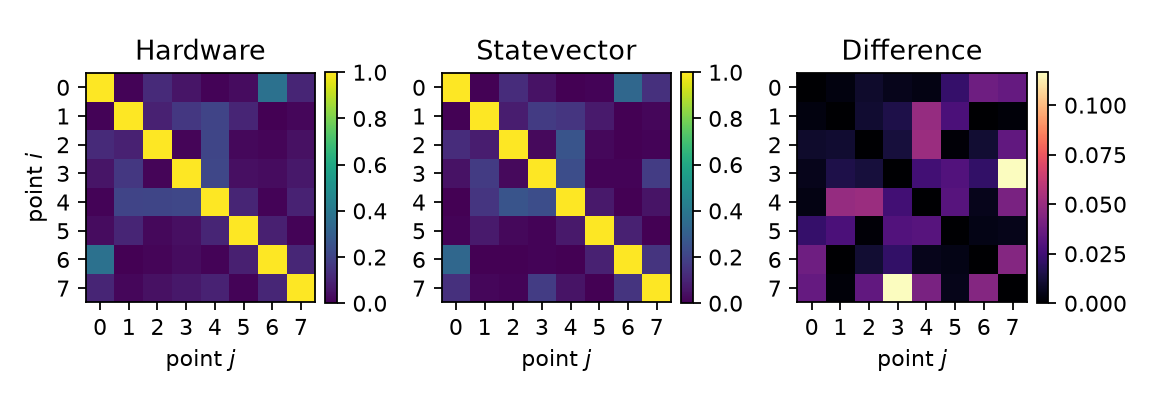}\\[0.8em]
\includegraphics[width=\textwidth]{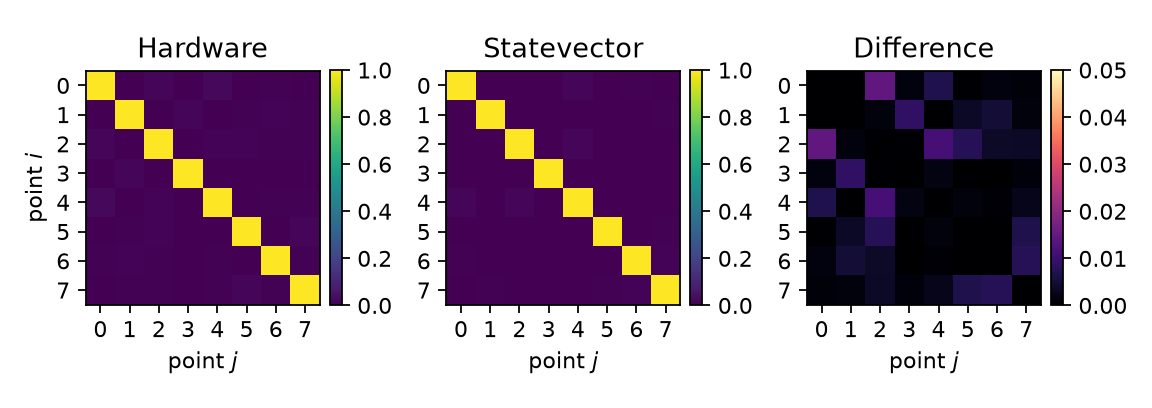}
\caption{Breast PCA fidelity kernels on \texttt{ibm\_fez}, $n=8$ points
($8\times 8$ matrices), $256$ shots.  Each row is the hardware kernel,
the statevector kernel, and their absolute difference.
\emph{Top}, $q=3=\lceil D_2\rceil$: off-diagonal structure is visible
and the hardware agrees with the statevector (MAE $0.021$).
\emph{Bottom}, $q=7$, past the ceiling: both kernels are essentially
the identity.  The algorithms of Table~\ref{tab:stack} would see eight
nearly orthogonal states.}
\label{fig:hwkernels}
\end{figure}

On breast, $q=3$ PCA is the geometrically structured kernel at $n=32$
in simulation; on device, with $n=8$, hardware still reproduces it
(MAE $0.021$; off-diagonals $0.082$ vs.\ $0.081$).  FD-ASE original
columns ($q=4$, indices $27,21,9,8$) match even more tightly (MAE
$0.012$).  A random triple ($15,18,23$) is not collapsed (mean
off-diagonal $0.16$) but $n=8$ cannot support the near/far alive
rule---the PCA $q=3$ subsample fails it too.  At $q=7$ the exact kernel
has already concentrated; hardware agrees (MAE $0.003$ because both
sides are near zero).  The ceiling is therefore a property of the
feature map, not of the simulator.  Blobs show the milder trend:
raising $q$ from $5$ to $7$ drops off-diagonals on both sides, with a
larger MAE from shot noise on a still-structured $K$.

\paragraph{\texorpdfstring{Selector comparison at $n=16$}%
  {Selector comparison at n=16}.}

The $n=8$ jobs show that the device reproduces the exact kernel, but
$\binom{8}{2}=28$ pairs are too few for the near/far diagnostic (only
a handful of pairs in each bin).  To test whether FD-ASE columns
outperform PCA and random columns \emph{on the device}, we run three
additional jobs with $n=16$ ($120$ compute--uncompute circuits each),
all on the same breast cancer rows: PCA at $q=3$, FD-ASE at $q=4$, and
three random columns at $q=3$.

Table~\ref{tab:hw16} reports the result.  The absolute fidelities are
low---the kernel is concentrated---so none of the three views passes the
full alive rule.  But the near/far \emph{ratio} discriminates clearly.
FD-ASE is the only view on which the hardware kernel still ranks
neighbours above strangers (ratio $2.1$ on device, $1.5$ on
statevector); PCA inverts the ordering (ratio $0.8$) and random
columns invert it more strongly (ratio $0.5$).  Hardware tracks the
simulator in every case (MAE $0.023$--$0.034$).

\begin{table}[t]
\centering
\caption{Selector comparison on hardware, $n=16$, one-layer $ZZ$.
FD-ASE is the only view where near $>$ far (ratio $>1$).}
\label{tab:hw16}
\small
\begin{tabular}{@{}lrrrrrr@{}}
\toprule
view & $q$ & MAE & near & far & ratio & mean $K$ \\
\midrule
PCA & 3 & 0.027 & 0.098 & 0.116 & 0.8 & 0.118 \\
FD-ASE & 4 & 0.023 & 0.106 & 0.050 & \textbf{2.1} & 0.063 \\
random & 3 & 0.034 & 0.084 & 0.162 & 0.5 & 0.133 \\
\bottomrule
\end{tabular}
\end{table}

The reading is that the fractal selector preserves the local geometry of
the data even after the QPU has estimated the kernel; PCA and random
columns do not.  This is the comparison that the $n=8$ jobs could not
make, and it supports the claim that FD-ASE is a better choice of
coordinates than PCA or random axes for the one-layer $ZZ$ kernel on
breast cancer data.

\section{Conclusion}
\label{sec:conclusion}

The question was whether fractal geometry helps \emph{encodings} in QML,
and whether that means ``use fewer points''.

\textbf{Fewer qubits, yes---for the entangled $ZZ$ layer used in
practice.}  Encoding at $q=\lceil D_2\rceil$ keeps that fidelity kernel
geometrically alive.  Encoding the PCA-95\% width or all $E$ attributes
feeds QSVM, fidelity $k$NN, spectral clustering, one-class SVM and KRR
a kernel that has already collapsed (breast: live at $3$, dead by
$4$--$10$; moons: live at $2$--$3$, PCA would encode $19$).  The forward selection of Section~\ref{sec:selector} names the original
columns that realise that width.
On the simulator, two attributes per qubit still work at the fractal
$q$; pouring PCA-95\% into those qubits does not.

\textbf{The integer is map-dependent, and bandwidth is part of the map.}
A second $ZZ$ layer is already dead at $\lceil D_2\rceil$.
Product-state $Z$ and linear IQP live past it (breast: last alive $6$
and $5$).  Shrinking the angle scale to $c=0.25$ buys extra live
qubits (breast knee $6$); stretching to $c=2$ kills the kernel at
every $q$ we tried.  We treat those as scope conditions, not as a
reason to ignore $D_2$ on the map and scale one actually runs.

\textbf{Fewer data points, no---that was not the test.}  $D_2$ is
estimated on up to $4000$ classical rows.  The QML kernel uses $n=32$
on the simulator as a common diagnostic, not because $32$ is optimal.
Wine and diabetes move from knees $2$ and $3$ at $n=32$ to $4$ and $5$
at $n=128$, still short of their fractal ceilings.  On IBM, $n=8$ is
enough to see that $q=3$ is implemented faithfully and $q=7$ is already
diagonal.

\textbf{What IBM adds.}  The ceiling of the one-layer $ZZ$ map is a
property of the feature map, not of the simulator: hardware matches the
live $K$ at the fractal width (MAE $0.012$--$0.021$) and the dead $K$
beyond it.  At $n=16$, where near/far can be scored, FD-ASE is the only
view on which the device kernel still ranks neighbours above strangers
(ratio $2.1$); PCA and random columns invert the ordering.

\textbf{The ablation is robust.}  Across four data sets with
$\lceil D_2\rceil$ ranging from $2$ (iris, moons) to $6$ (pendigits),
the one-layer $ZZ$ knee at $c=1$ tracks the fractal ceiling
(Table~\ref{tab:ablation}).  Ten kernel samples put the breast and
moons medians at $q=3$.  A synthetic $k$-sweep shows that $D_2$
recovers known width on an axis-aligned cube, while the $ZZ$ map at
$c=1$ still dies near three full-range coordinates when $k$ is large.
After a Haar rotation of that cube, PCA-$95\%$ still equals $k$ and
the box-count ceiling does not (Section~\ref{sec:oblique}).  The budget is a
property of the map--data--bandwidth triple, and for the entangled
$ZZ$ layer used in practice at $c=1$ it is tight on low-ID tables.

\paragraph{Limitations.}
Box-count $D_2$ needs sample size and several scales; a few dozen points
are not enough.  FD-ASE selects original columns.  On a rotated linear subspace the
box-count ceiling and FD-ASE both overshoot $k$, and PCA-$95\%$ is the
selector that still equals $k$ (Section~\ref{sec:oblique}).  The alive rule is operational, not a
hypothesis test.  Effective rank, which uses no floors, rises toward
$n$ on the same sweeps and is already above $25$ at the first dead
$q$ (Section~\ref{sec:erank}); the integer compared with
$\lceil D_2\rceil$ remains the alive knee.  At $n=128$ that alive bit is
stable across ten kernel samples, but a dead kernel does not force
every downstream score down: breast fidelity $k$NN is unchanged from
$q=3$ to $q=8$, and breast SVM F1 at $q=8$ ranges from $0$ to $0.92$
(Section~\ref{sec:downstream}).  Amplitude encoding ($\lceil\log_2 E\rceil$ qubits)
and projected kernels \citep{huang2021power,agliardi2025mitigating}
change \emph{how} $K$ is evaluated, not \emph{which} coordinates enter
the map.  The fractal ceiling is tight for one-layer $ZZ$ at $c=1$ and loose for
product-state $Z$ or for $c\ll 1$; quoting $\lceil D_2\rceil$ without
naming the map and the bandwidth would overclaim.  Hardware here is
$n=8$ and $n=16$, one backend, $256$ shots, no error mitigation, and
only the map whose simulator knee tracks $D_2$.  The $n=16$ selector
comparison is on one data set (breast).  FD-ASE is not alive on every
table at $n=32$; it is a column selector, not a second ceiling.

\backmatter

\section*{Statements and Declarations}

\noindent\textbf{Funding.} Not applicable.

\noindent\textbf{Competing interests.} The author declares no competing interests.

\noindent\textbf{Ethics approval and consent to participate.} Not applicable.

\noindent\textbf{Consent for publication.} Not applicable.

\noindent\textbf{Data availability.}
Iris, wine, and the Wisconsin diagnostic breast-cancer table are the
scikit-learn copies of UCI sets 53, 109 and 17.  Digits are the
scikit-learn copy of UCI Optical Recognition of Handwritten Digits
(set 80).  Pendigits is the UCI pen-based set (set 81), training and
test files together.  Diabetes is the scikit-learn copy of the baseline
table in \citet{efron2004least}, not the UCI Pima set.  Moons are
\texttt{sklearn.datasets.make\_moons} with eighteen noise columns.
Intrinsic2 and OneBig are generated in the code repository.
The UCI addresses are
\begin{flushleft}\small
\url{https://archive.ics.uci.edu/dataset/53}\\
\url{https://archive.ics.uci.edu/dataset/109}\\
\url{https://archive.ics.uci.edu/dataset/17}\\
\url{https://archive.ics.uci.edu/dataset/80}\\
\url{https://archive.ics.uci.edu/dataset/81}
\end{flushleft}
Statevector sweep tables, figures, and the IBM Quantum kernel matrices
are archived with the code.

\noindent\textbf{Materials availability.} Not applicable.

\noindent\textbf{Code availability.} The code that estimates $D_2$, runs FD-ASE, and builds the fidelity kernels is available at \url{https://github.com/anapaulaappel/qml-qubit-budget}.

\noindent\textbf{Author contribution.} The author conceived the study, carried out the experiments, and wrote the manuscript.

\bibliography{referencias}

\end{document}